# Ab-initio investigation of the physical properties of BaAgAs Dirac semimetal and its possible thermo-mechanical and optoelectronic applications

A.S.M. Muhasin Reza, S.H. Naqib*

Department of Physics, University of Rajshahi, Rajshahi 6205, Bangladesh

*Corresponding author; email: salehnaqib@yahoo.com

**Abstract**

BaAgAs is a ternary Dirac semimetal which can be tuned across a number of topological orders. In this study we have investigated the bulk physical properties of BaAgAs using density functional theory based computations. Most of the results presented in this work are novel. The optimized structural parameters are in good agreement with previous results. The elastic constants indicate that BaAgAs is mechanically stable and brittle in nature. The compound is moderately hard and possesses fair degree of machinability. There is significant mechanical/elastic anisotropy in BaAgAs. The Debye temperature of the compound is medium and the phonon thermal conductivity and melting temperature are moderate as well. The bonding character is mixed with notable covalent contribution. The electronic band structure calculations reveal clear semimetallic behavior with a Dirac node at the Fermi level. BaAgAs has a small ellipsoidal Fermi surface centered at the G-point of the Brillouin zone. The phonon dispersion curves show dynamical stability. There is a clear phonon band gap between the acoustic and the optical branches. The energy dependent optical constants conform to the band structure calculations. The compound is an efficient absorber of the ultraviolet light and has potential to be used as an anti-reflection coating. Optical anisotropy of BaAgAs is moderate. The computed repulsive Coulomb pseudopotential is low indicating that the electronic correlations in this compound are not strong.

## 1. Introduction

Experimental and theoretical studies of topological semimetals have become a major branch in the condensed matter physics research. These compounds are characterized by semimetallic electronic band structures with non-trivial band crossings which are protected by symmetries. Topological semimetals, with Dirac, Weyl, Nodal lines bulk band topological signatures [1-9]. Recently, a hexagonal ternary compound BaAgAs has been characterized experimentally [10] which crystallizes in the space group $P6_3/mmc$ (No.194) [11,12]. First-principles calculations suggested that BaAgAs is a Dirac semimetal (DSM) with a pair of Dirac points lying on the $C_3$ rotation axis. Moreover potassium doping in BaAgAs can transform the system into a triple-point semimetallic (TPSM) state [11-14].

A few other compounds having the same class of BaAgAs were also experimentally synthesized earlier, e.g., BaAgBi, SrCuBi, and SrAgBi. These compounds also show perfect DSM phase at ambient conditions [11,12]. There is significant interest in DSM compounds because of their high charge carrier mobility, low carrier concentration, large magnetoresistance (MR) and chiral anomaly-induced negative MR. All these features are useful in electronics and spintronics device applications. In topological semimetals, specific states of electrons are topologically protected and are free from environmental perturbation.

According to first-principles calculations, when the spin-orbit coupling (SOC) is ignored, the electronic band structure of BaAgAs hosts a broken-symmetry-driven topological state [15]. In contrast, when the SOC is incorporated, the nodal crossings become protected by the double point group representations at k points along the $C_3$ axis [10]. As far as the structure of BaAgAs in concerned, the planar mono-hexagonal layer is loosely occupied by the Ba atoms, while the planar honeycomb lattice is constructed by two different hexagonal sublattices that are occupied by the Ag and As atoms, alternately. For thermal transport studies, a low Grüneisen parameter is suggested by the phonon calculations. The BaAgAs compound with high lattice symmetry attains low thermal lattice conductivity due to the heavy element Ba in planar mono-hexagonal layer and large mass fluctuations in the Ag–As planar honeycomb sublattices [16]. The electronic properties of topological semimetals, like the gapless band structure can be used in photo detectors [17–19]. Fermi arc states for spintronics and structural materials [20,21] and qubits [22,23] and for nanoscale device applications in nanostructures. In recent studies the electronic band structure and magneto-transport properties of BaAgAs have been studied in some details [24,25].

To the best of our knowledge, there are no available experimental or theoretical studies on the bulk elastic, mechanical, acoustic, bonding, optical, lattice dynamical and thermo-mechanical properties of BaAgAs yet. All these unexplored bulk properties are important to understand this compound better and to unlock its potential for applications in engineering and optoelectronic sectors. We aim to fill the significant research gap existing for this topological semimetal. We have calculated the elastic constants and moduli for the optimized crystal structure of BaAgAs. The important mechanical performance indicators like hardness and machinability index have been evaluated. The elastic/mechanical anisotropy parameters are estimated. The nature of chemical bonding has been reported. The electronic band structure and the Fermi surface topology have been revisited. The phonon dispersion curves have been calculated and the optical properties have been explored in details. Some thermo-mechanical parameters, pertinent to applications of BaAgAs are computed. Most of the results presented in this study are entirely novel.

The rest of the paper has been arranged as follows. The computational methodology is described in Section 2. The results of the calculations are presented and discussed in Section 3. Section 4 consists of the conclusions of this work.

**2. Computational scheme**

In this work all the calculations are performed by using the plane wave pseudopotential density functional theory (DFT) method as contained in the CAmbridge Serial Total Energy Package (CASTEP) [26–28]. The exchange-correlation terms are incorporated in the total energy by using the local density approximation (LDA) with the functional CA-PZ. The ground state of the crystalline solid is found from the solution of the Kohn-Sham equation [27]. For reliable results, selection of the atomic core pseudopotential is important. The pseudopotential gives the residual attractive interaction between an electron and an ion after taking into account the effective repulsion that arises from the exclusion principle demanding that valence states are orthogonal to the core electronic states. The on-the-fly generated (OTFG) ultrasoft pseudopotential has been used in the calculations [26-28]. The Broyden-Fletcher-Goldfarb-Shanno (BFGS) optimization method has been adopted to find out the ground state crystal structure. We have also used the density mixing [29]. The following valence electron orbitals are considered for Ba, Ag and As atoms, respectively: Ba [$6s^26p^6$], Ag [$4d^{10}5s^1$] and As [$4s^24p^3$]. The Γ-centered k-points have been considered in the reciprocal space (Brillouin zone). The convergence criteria for structure optimization and energy calculations were set to ultrafine quality with the k-points mesh of size 15×15×7 in the Monkhorst-Pack grid scheme [30] has been used for the sampling of the first Brilloun zone (BZ) of the hexagonal unit cell of BaAgAs. A plane wave basis with a cut off energy of 500 eV is used to expand the eigenfunctions of the valence and nearly valence electrons. Geometry optimization has been performed using self-consistent convergence limits of 5×10⁻⁶ eV/atom for the energy, 0.01 eV/Å for the maximum force, 0.02 GPa for the maximum stress and 5×10⁻⁴ Å for maximum atomic displacement.

The optical properties of BaAgAs have been evaluated using the electronic band structure for the optimum crystal structure. Further details regarding calculations of the optical parameters' spectra can be found in Refs. [26-28]. The single crystal elastic constants are obtained using the stress-strain method contained in the CASTEP. Thermal parameters have been studied with the help of various elastic constants and moduli. The phonon dispersion calculations are carried out using the perturbative linear response theory. In the electronic band structure calculations, we have not included the spin-orbit coupling (SOC). From a number of previous studies, we have found that the bulk physical properties of topological semimetals are fairly insensitive to the SOC, particularly for compounds where SOC are not very strong [31-34]. Inclusion of the SOC affects mainly the surface electronic states and some of the bulk electronic bands get splitted in energy. Previous electronic band structures for BaAgAs with and without SOC show that the energy splitting of the electronic bands are not that significant [10].

The chemical bonding natures of BaAgAs have been explored via the Mulliken population analysis (MPA) and the Hirshfeld population analysis (HPA). The details regarding MPA and HPA can be found elsewhere [26-28].

**3. Results and analysis**

**3.1. Structural properties**

As mentioned earlier, the crystal structure of BaAgAs is hexagonal with space group P6₃/mmc (No. 194). The schematic crystal structure of BaAgAs is shown in Figure 1. It is clear from the crystal structure that BaAgAs consists of alternative layers of Ba and AgAs along the c-axis. There are two AgAs layers in one unit cell. The AgAs layers form triangular lattices and are sandwiched between trigonal Ba layers along the c-axis [10,12]. The unit cell consists of six atoms in which there are two Ba atoms, two Ag atoms and two As atoms. The atomic positions and lattice parameters of the crystal are fully relaxed starting with the experimental values found in earlier studies (Table 1). The optimized lattice constants a (= b) and c obtained using the LDA calculations along with experimental lattice constants and other theoretical values are listed in Table 1. The positions of atoms in BaAgAs are as follows: Ba atoms are placed at the positions (0, 0, 0), the Ag atoms are at (1/3, 2/3, 3/4) and the As atoms are at (1/3, 2/3, 1/4) [10,12]. It is observed that the present values are close to the experimental ones [12,25]. Since optimization of the crystal geometry is one of the most crucial part in any ab-initio investigation, fair agreement between the computed and experimental lattice constants imply that the results obtained in this study are reliable [12,15,25].

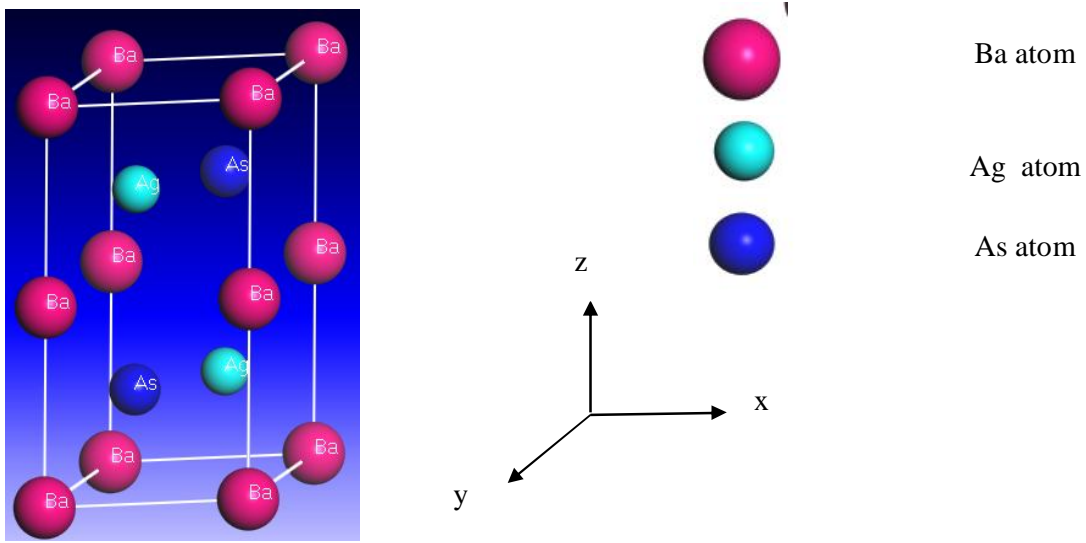

**Figure 1:** Schematic crystal structure of BaAgAs. The crystallographic directions are shown and different atoms are presented in different colors.

**Table 1:** Calculated lattice constants a (= b) and c, c/a ratio, and equilibrium cell volume of hexagonal BaAgAs.

| Compound | a = b (Å) | c (Å) | c/a | Volume $V_0$ (Å$^3$)* | Ref. |
|---|---|---|---|---|---|
| BaAgAs | 4.377 | 7.939 | 1.813 | 131.72 | [25]$^{Theo.}$ |
|  | 4.450 | 8.040 | 1.806 | 137.90 | [25]$^{Exp.}$ |
|  | 4.561 | 8.419 | 1.845 | 151.67 | [15]$^{Theo.}$ |
|  | 4.521 | 8.281 | 1.832 | 147.02 | [15]$^{Exp.}$ |
|  | 4.496 | 8.828 | 1.963 | 158.41 | [12]$^{Exp}$ |
|  | 4.484 | 8.819 | 1.967 | 153.65 | This work |

*Unit cell volume = (abc)sin60$^0$

It is seen from Table 1 that there are some scatter in the values of the lattice parameters obtained by different groups. As far as theoretical values are concerned, this is due to the use of different exchange-correlation functionals and/or computational set-ups. Generally, LDA underestimates the lattice constants due to overbinding of the atoms [26]. It is also instructive to note that the experimental values are often obtained at room temperature while the theoretical values are for the optimized crystal structure in the ground state (zero degree Kelvin).

### 3.2. Elastic properties

#### 3.2.1. The stiffness constants

The mechanical properties of crystalline solids are determined by the single crystal elastic constants ($C_{ij}$). The possible mechanical applications of solids are limited by their elastic behavior. The elastic constants are connected with the structural features and atomic bonding characteristics of materials. The elastic constants also determine the mechanical stability of a solid. The bulk elastic behavior is understood from the polycrystalline elastic moduli. The compound under study is hexagonal and therefore, it has five independent elastic constants: $C_{11}$, $C_{12}$, $C_{13}$, $C_{33}$, and $C_{44}$. The elastic constant $C_{66}$ is not independent as it can be expressed as: $C_{66} = \frac{C_{11} - C_{12}}{2}$. The mechanical stability conditions of a hexagonal crystal system are as follows [36,37].

$$C_{11} > 0;\ C_{11} > C_{12};\ C_{44} > 0;\ (C_{11}+C_{12})C_{33} - 2(C_{13})^2 > 0 \qquad (1)$$

All these conditions are satisfied by the computed $C_{ij}$s of BaAgAs, and hence the compound under study is expected to be mechanically stable. We have also calculated the tetragonal shear modulus, given by, $(C_{11} - C_{12})/2$, for BaAgAs. The tetragonal shear modulus corresponds to a specific phonon vibration mode and is thus directional in nature. A positive value of this parameter indicates dynamical stability.

Among the five independent elastic constants, $C_{11}$ and $C_{33}$ are the measures of stiffness along the a- and c-axes of the crystal, respectively. From Table 2 it is seen that $C_{11} > C_{33}$, so the atomic bonding is much stronger along the a-axis than that along the c-axis of BaAgAs. Since both the elastic constants $C_{11}$ and $C_{33}$ are significantly larger than $C_{44}$, the linear compression along the crystallographic a- and c-axes is rather difficult in comparison with the shear deformation. The other three shear related elastic constants have values close to $C_{44}$. The elastic constant $C_{44}$ is linked to the hardness of a crystal and machinability index [38,39]. Since no reported values of $C_{ij}$ are available for BaAgAs, we have compared the computed results with those of another isostructural Dirac semimetal BaAgP in Table 2. It is seen that all the elastic constants of BaAgAs are larger than those of BaAgP. Therefore, we expect that BaAgAs is harder than BaAgP.

**Table 2:** The calculated elastic constants, $C_{ij}$ (GPa), of BaAgAs in the ground state.

| Compound | $C_{11} (= C_{22})$ | $C_{12}$ | $C_{13} (= C_{23})$ | $C_{33}$ | $C_{44} (= C_{55})$ | $C_{66}$ | Ref. |
|---|---|---|---|---|---|---|---|
| BaAgAs | 131.86 | 43.61 | 35.08 | 76.35 | 33.87 | 44.12 | This work |
| BaAgP | 115.80 | 31.10 | 22.80 | 71.30 | 30.40 | 42.30 | [40] |

*3.2.2. Elastic moduli and parameters*

To calculate the polycrystalline elastic moduli B (bulk modulus), G (shear modulus), Y (Young's modulus), we have used the Voigt-Reuss-Hill formalism [41–43]. Further information on the widely used equations connecting the elastic moduli to the single crystal elastic constants can found elsewhere [44-46]. The results obtained are listed in Table 3. Smaller value of G compared to B (Table 3) indicates that the mechanical stability of BaAgAs will be controlled by the shear modulus. From the ratio between tensile stress and strain we get the Young's modulus, which is the measurement of the resistance (stiffness) of an elastic solid to a change in its length and provides with a measure of thermal shock resistance. There are other useful bulk elastic parameters, e.g., the Pugh's ratio (k), Poisson's ratio ($\sigma$), and the machinability index ($\mu_M$). All these parameters can be calculated using the following expressions:

$$k = \frac{B}{G} \qquad (2)$$

$$\sigma = \frac{3B - 2G}{6B + 2G} \qquad (3)$$

$$\mu_M = \frac{B}{C_{44}} \qquad (4)$$

The bulk modulus B shows the resistance to fracture and shear modulus G represents resistance to plastic deformation. The macro hardness ($H_{macro}$) and the micro hardness ($H_{micro}$) parameters of BaAgAs are also calculated using the following formulae [47,48]:

$$H_{macro} = 2[(\frac{G}{B})^2 G]^{0.585} - 3 \quad (5)$$

$$H_{micro} = \frac{(1-2\sigma)Y}{6(1+\sigma)} \quad (6)$$

The Cauchy pressure, $C_p$, is another important elastic parameter closely related to the bonding character and brittleness/ductility of crystalline solids. For hexagonal systems, $C_p = (C_{12} - C_{44})$. The computed elastic moduli and other elastic parameters are summarized in Table 3 below. In the absence of prior elastic data for BaAgAs, we have compared the values with those of BaAgP.

**Table 3:** The computed bulk modulus B (GPa), shear modulus G (GPa), Young's modulus Y (GPa), machinability index $\mu_M$, macro hardness $H_{macro}$ (GPa), micro hardness, $H_{micro}$ (GPa), Pugh's ratio k (= B/G), Poisson's ratio σ and the Cauchy pressure $C_p$ (GPa) of BaAgAs.

| Compound | B | G | Y | $\mu_m$ | $H_{macro}$ | $H_{micro}$ | k | σ | $C_p$ | Ref. |
|---|---|---|---|---|---|---|---|---|---|---|
| BaAgAs | 60.64 | 36.72 | 91.65 | 1.79 | 6.05 | 6.16 | 1.65 | 0.25 | 9.74 | This work |
| BaAgP | 49.20 | 35.00 | 85.00 | 1.62 | - | - | 1.41 | 0.21 | 0.70 | [40] |

The machinability index measures the level of plasticity and dry lubricity of a solid [49-51]. It also gives indication about the ease with which a solid can be cut into desired shapes. The machinability index of BaAgAs is high, comparable to many ternary MAX and MAB phase compounds which are promising materials for engineering structural applications [52-55]. This implies that the compound under consideration is machinable and has significant dry lubricity. Both the macro and micro hardnesses are moderate suggesting that the overall bonding strength in BaAgAs is not very strong. The Pugh's ratio determines the failure modes of materials. Solids having a Pugh's ratio greater than 1.75 are ductile in nature. If the Pugh's ratio is below this value, then the material is predicted to be brittle in nature [56]. The obtained value of Pugh's ratio of BaAgAs is 1.65 which is less than 1.75. Therefore, BaAgAs should show brittleness. The nature of bonding can be guessed from the value of the Poisson's ratio. The central force interactions dominate in solids when the value of Poisson's ratio lies between 0.25 to 0.50. For BaAgAs, the value of Poisson's ratio is 0.25 meaning that the chemical bonding has central force nature. Moreover, for a purely covalent crystal the value of Poisson's ratio is around 0.10 and for a crystal with dominant ionic/metallic bonding, the value of Poisson's ratio is around 0.33 [57]. Therefore, we expect some contribution of ionic/metallic bondings in BaAgAs. The value of the Poisson's ratio can also be used to determine brittleness/ductility. If σ is less than the critical value 0.26, then a material is predicted to be brittle; it will be ductile otherwise. The computed Poisson's ratio is 0.25 for BaAgAs indicating the brittle nature [58]. The Cauchy pressure can also be used to separate the materials into brittle or ductile category where the critical value of $C_p$ is zero [59]. The obtained value of $C_p$ of BaAgAs is positive; signifying that it is our material is ductile in nature. This contradicts the findings from the Pugh's ratio and the Poisson's ratio. The contradiction can arise from quantum many-body interactions among the atoms which are omitted in the formalism for Cauchy pressure and can lead to sign problem in solids which are situated at the borderline of ductile/brittle behavior. According to Pettifor's rule [60], materials with positive Cauchy pressure have metallic bonds. Negative Cauchy pressure, on the other hand, suggests that the presence of angular bondings. The results presented in Table 3 are novel and therefore, no comparison with any other reported values can be made. Elastic moduli of BaAgAs are larger than those of BaAgP. BaAgAs is more machinable than the BaAgP compound. Both the compounds are expected to be moderately brittle in nature.

### 3.2.3. Elastic anisotropy

Most of the crystalline solids are elastically anisotropic. Elastic/mechanical anisotropy depends on the atomic arrangements within the unit cell. Elastic anisotropy is connected to many important mechanical properties of solids which are related to their practical applications [21]. The elastic anisotropy of BaAgAs has been studied by means of different anisotropy indices. The Zener anisotropy factor (A) has been calculated using the relation [61].

$$A = \frac{2C_{44}}{(C_{11} - C_{12})} \tag{7}$$

This parameter quantifies the level of anisotropy in the elastic constants in a single crystal. For an isotropic crystal, A = 1. Deviation from unity gives the measure of anisotropy. For BaAgAs, A = 0.76, indicating that the compound is anisotropic.

The shear anisotropy factors measure the anisotropy in bonding strengths among atoms situated in different crystal planes. There are three different shear anisotropy factors for hexagonal crystal. These factors can be computed from the following equations [61,62]:

$$A_1 = \frac{(C_{11} + C_{12} + 2C_{33} - 4C_{13})}{6C_{44}} \tag{8}$$

for {100} shear planes between <011> and <010> directions.

$$A_2 = \frac{2C_{44}}{C_{11} - C_{12}} \tag{9}$$

for {010} shear planes between < 101> and <001> directions.

$$A_3 = \frac{C_{11} + C_{12} + 2C_{33} - 4C_{13}}{3(C_{11} - C_{12})} \tag{10}$$

for {001} shear planes between <110> and <010> directions.

These factors, $A_i$ (i = 1, 2, 3), have the unit value for shear-isotropic crystals. Departure from unity quantifies the level of anisotropy in the shape changing deformation due to shearing stresses on different crystal planes. Calculated values are disclosed in Table 4.

The directional bulk modulus along the a-direction and c-direction can be estimated by using the following relations [62]:

$$B_a = \alpha \frac{dP}{da} = \frac{\Lambda}{(2 + a)} \tag{11}$$

$$B_c = \alpha \frac{dP}{dc} = \frac{B_a}{a} \tag{12}$$

where, $\Lambda = 2(C_{11} + C_{12}) + 4C_{13}\alpha + C_{33}\alpha^2$ \hfill (13)

and $\alpha = \frac{(C_{11} + C_{12} - 2C_{13})}{(C_{33} + C_{13})}$ \hfill (14)

The linear compressibility along the a-axis ($\chi_a$) and c-axis ($\chi_c$) can be calculated using the following relations [62]:

$$\chi_a = -\frac{1}{a}\left(\frac{\partial a}{\partial p}\right) = \frac{(C_{33} - C_{13})}{C_{33}(C_{11} + C_{12}) - 2C_{13}^2} \quad (15)$$

$$\chi_c = -\frac{1}{c}\left(\frac{\partial c}{\partial p}\right) = \frac{(C_{11} + C_{12} - 2C_{13})}{C_{33}(C_{11} + C_{12}) - 2C_{13}^2} \quad (16)$$

The calculated values of $\chi_a$ and $\chi_c$ for BaAgAs are $3.77 \times 10^{-3}$ GPa$^{-1}$ and $9.62 \times 10^{-3}$ GPa$^{-1}$, respectively. These values indicate that the compressibility along the c-axis is more than double that of along a-axis. This is a clear indication that the atomic bonding strengths are much weaker along the c-axis. The ratio of the two linear compressibility coefficients along the a- and c- axes of hexagonal crystals, $\frac{c_c}{c_a}$ is another useful parameter to understand the in-plane and out-of-plane anisotropy in the bonding strengths [61,62]:

$$\frac{c_c}{c_a} = \frac{(C_{11} + C_{12} - 2C_{13})}{(C_{33} - C_{13})} \quad (17)$$

All these anisotropy factors are evaluated and listed in Table 4. It should be noted that $B_a = B_c$ for isotropic crystals and and $\frac{c_c}{c_a} = 1$, for the same. In the absence of any prior work on the elastic anisotropy of BaAgAs, we have presented some anisotropy results for isostructural BaAgP Dirac semimetal in Table 4 for comparison.

**Table 4:** Zener anisotropy factor (A), Shear anisotropy factors ($A_1$, $A_2$ and $A_3$), directional bulk moduli ($B_a$, $B_c$ in GPa), linear compressibility coefficients ($\chi_a$, $\chi_c$ in $10^{-3}$ GPa$^{-1}$) and ratio of the linear compressibility coefficients $\frac{c_c}{c_a}$ of BaAgAs.

| Compound | A | $A_1$ | $A_2$ | $A_3$ | $B_a$ | $B_c$ | $\chi_a$ | $\chi_c$ | $\frac{c_c}{c_a}$ | Ref. |
|---|---|---|---|---|---|---|---|---|---|---|
| BaAgAs | 0.76 | 0.92 | 0.76 | 0.70 | 187.34 | 198.24 | 3.77 | 9.62 | 2.55 | This work |
| BaAgP | 0.71 | - | - | - | - | - | 5.14 | 10.74 | 2.08 | [40] |

Table 4 reveals that the shear anisotropy is significant in BaAgAs. The anisotropy in the bulk moduli is moderate. The level of elastic anisotropy in BaAgAs and BaAgP are roughly similar.

### 3.4. Electronic Properties

#### *3.4.1. Electronic band structure*

Electronic band structure is one of the most important aspects of materials which control all the electronic and optical properties. It also gives information regarding atomic bonding in a crystal and stability of a material [63]. The bulk electronic band structure as a function of energy (E-E$_F$) along the high symmetry directions in the Brillouin zone (BZ) is calculated in the ground state and is shown in Figure 2. The Fermi level (E$_F$) is indicated by the horizontal broken line which has been set at zero eV. The compound is semi-metallic due to weak crossing of the energy band of the Fermi level around the G-point. This weak crossing implies that the compound has small Fermi sheet centered at the G-point. The Dirac node touching the Fermi level at the G-point (along the K-G-M line) conforms the topological electronic state. The conduction bands show both electron- and hole-like energy dispersions. The bands close to the Fermi level are due to the Ba-5s, Ag-4s, Ag-4p and As-4s, As-4p electronic orbitals.

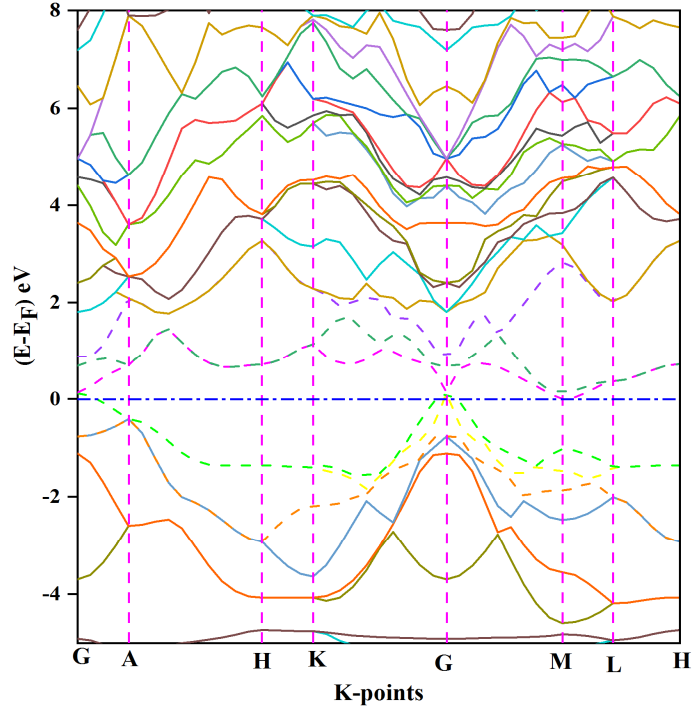

**Figure 2:** Electronic band structure of BaAgAs in the ground state. The dashed horizontal line marks the Fermi energy (set to 0 eV).

*3.4.2. Electronic energy density of states (DOS)*

The electronic energy density of states (DOS) is defined as the number of available electronic states per unit energy per unit volume. It is related to the first derivative of the energy dispersion curves E(k) or band structures with respect to momentum (k). DOS is inversely proportional to this derivative, i.e., the DOS is small where the derivative is large and vice versa. On the other hand, the effective mass of electrons or holes is directly proportional to the inverse of the second derivative of E(k). Large number of charge transport, optoelectronic, and magnetic properties of materials are directly determined by the DOS close to the Fermi energy.

In this section, the total and partial densities of states (PDOS and TDOS, respectively) of BaAgAs are calculated from the electronic band structure results. The PDOS and TDOS plots are given in Fig. 3. The vertical broken line indicates the Fermi level. The non-zero value of TDOS at the Fermi level confirms that BaAgAs will exhibit metallic electrical conductivity. To investigate the contribution of each atom to the TDOS of BaAgAs, we have shown the PDOS of electrons in Ba, Ag and As atoms separately. The TDOS value at the Fermi level is 0.667 states/eV-formula unit. This small value of the TDOS suggests semimetallic character of BaAgAs. The large peaks in the TDOS in the valence band centered at -1.47 eV and at 3.94 eV in the conduction band are principally responsible for the charge transport and optoelectronic properties of BaAgAs. These two peaks are due to the Ag-4s, Ag-4d, As-4p and Ba-4d electronic states. The overall contribution of the electronic states of the As atom in the energy range shown in Fig. 3 is quite small. The Fermi level is located quite close to the pseudogap separating the bonding and the antibonding peaks. This suggests that BaAgAs has high structural stability.

One can estimate the degree of electronic correlation in BaAgAs using the TDOS at the Fermi level, $N(E_F)$. The repulsive Coulomb pseudopotential, $V_c$, is a measure of the electronic correlation which is related to $N(E_F)$ as follows [64].

$$V_c = 0.26 N(E_F)/[1+N(E_F)] \qquad (18)$$

The calculated value of $V_c$ turns out to be 0.104. This shows that electronic correlation is not strong in BaAgAs.

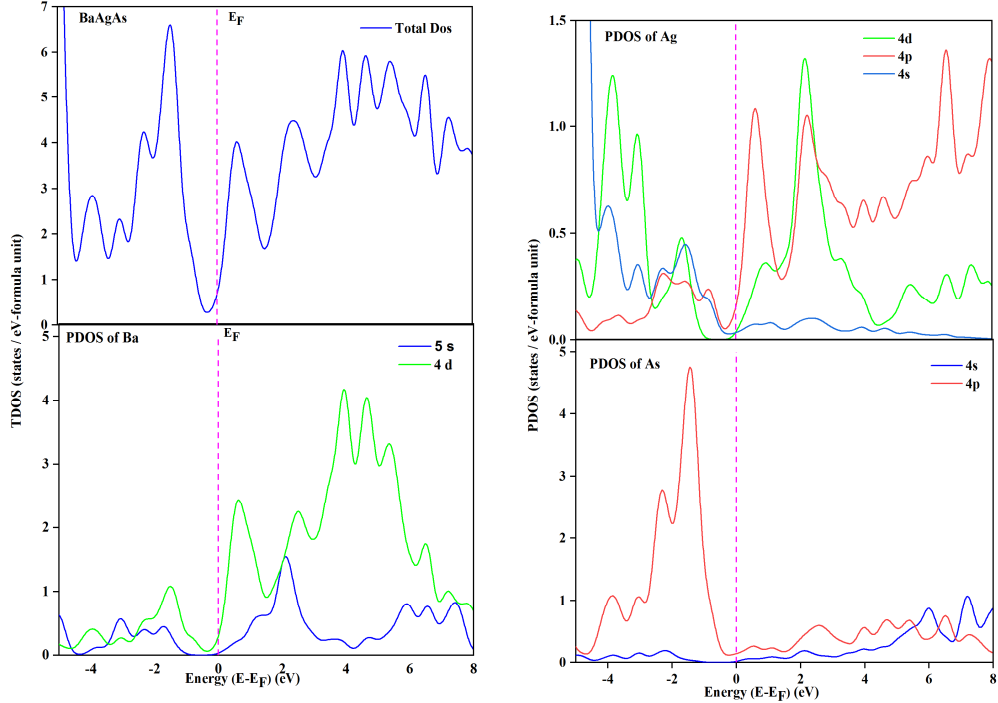

**Figure 3:** Total and partial electronic densities of states (TDOS and PDOS) of BaAgAs in the ground state.

### *3.4.3. Electronic charge density distribution*

The charge distribution around each atom in the crystal is very important to understand the bonding nature. In this section we have studied the electronic charge density distribution in different crystal planes of BaAgAs. The electronic charge density distribution of BaAgAs in the (111) and (001) planes are given in Fig. 4. The color scale shown on the right hand side of the panels represents the total electron density. The blue color indicates the high charge (electron) density and the red color indicates the low charge (electron) density. The charge density maps show mixed character of chemical bondings. In the (111) plane, there is significant directional accumulation of electronic charge in the Ag atom. The same behavior is found for the As atom in the (001) plane. The charged contours around the Ag and As atoms are not completely circular, indicating that both ionic and covalent contributions are present. There is charge depletion in between As and Ag atoms in the (001) plane. The charge distribution around the Ba atom located in between the As and Ag atoms are severely distorted and shows strong directionality in the (001) plane. Similar distortion is also seen for the Ba atom in the (111) plane. The charge accumulation in between the As and Ag atoms in the (111) plane are indicative of weak covalent bonding. From the charge density maps of BaAgAs in both the planes, we can see that Ag and As atoms have high electron density compared to the Ba atoms. The low charge concentration for the Ba atoms implies that the uniform background charges (the red region) probably come primarily from the Ba electrons in the conduction band.

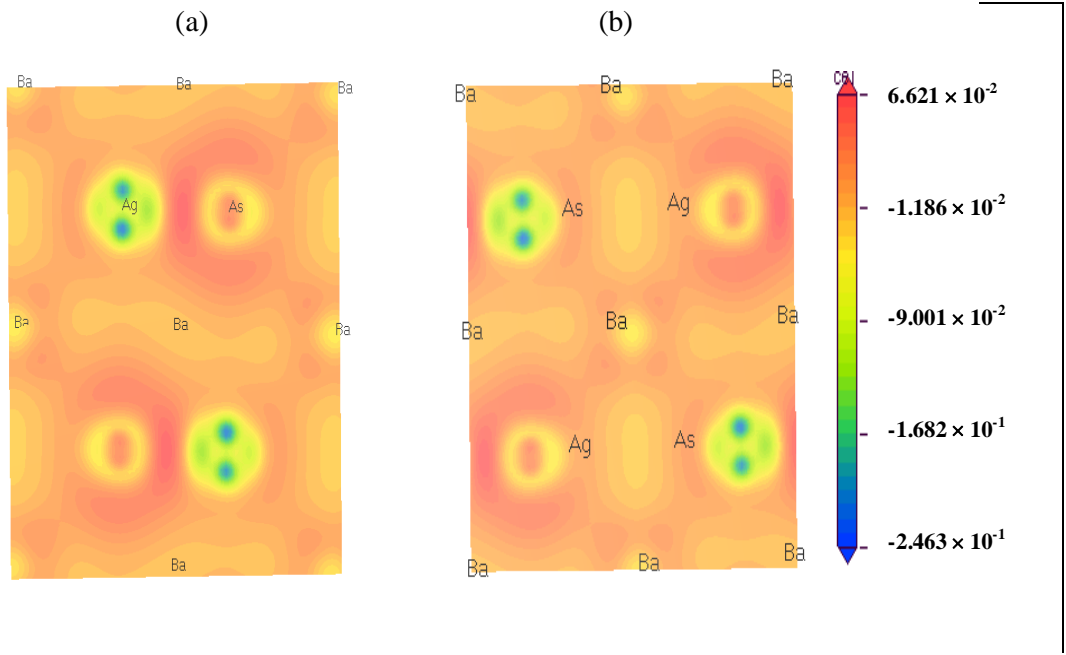

**Figure 4:** The electronic charge density distribution map for BaAgAs in the (a) (111) and (b) (001) planes. The color scale on the left quantifies the amount of charge (in unit of electronic charge).

*3.4.4. Fermi surface*

The Fermi surface of BaAgAs is shown in Fig. 5. From the band structure of BaAgAs we have found just one band crossing the Fermi level. This band contributes to the single sections of the Fermi surface centered on the G-point in the reciprocal lattice. Weak crossing leads to a small Fermi surface and implies that the electrical and electronic thermal conductivity of the Dirac semimetal BaAgAs should be quite low. The Fermi sheet enclosing the central ellipsoid has electronic character. The shape of the ellipsoid suggests that the energy dispersion is much stronger in the ab-plane compared to that along the c-axis. This in turn implies that there is anisotropy in the effective mass of the electrons traveling in the ab-plane and perpendicular to the ab-plane.

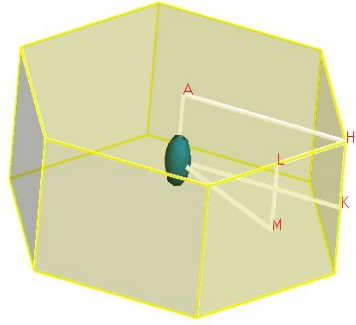

**Figure 5:** Fermi Surface of BaAgAs. The symmetry directions in the Brillouin zone are shown.

### 3.5. Phonon dynamics and acoustic properties

*3.5.1. Phonon dispersion curves and phonon density of states*

The characteristics of phonons are important for the understanding of physical properties of crystalline materials. Phonons are the energy quanta of lattice vibrations. Large number of electrical, thermal, elastic, and lattice dynamical properties of crystalline solids are dependent on the phonon spectrum. The phonon spectrum is determined by the crystal symmetry, stiffness constants and mass of the constituent atoms [65–67]. The electron–phonon interaction function is directly related to the phonon density of states (DOS). From phonon dispersion spectra (PDS) we can determine the structural stability, indication of possible structural phase transition and thermal properties of a solid. By using the linear perturbative approach [28], we have calculated the phonon dispersion spectra (PDS) and phonon density of states (PHDOS) of BaAgAs compound along the high symmetry directions of the first Brillouin zone which have been shown in Fig. 6. Since all the phonon modes within the first BZ are positive for BaAgAs, the compound is dynamically stable. The speed of propagation of an acoustic phonon, which is also the sound speed in the lattice, is given by the slope of the acoustic dispersion relation, $\partial\omega/\partial k$. At low values of k (i.e. in the long wavelength limit), the dispersion relation is almost linear, independent of the phonon frequency. This behavior does not hold at large values of k, i.e. for short wavelengths. These short wavelength phonon modes are generally the optical modes.

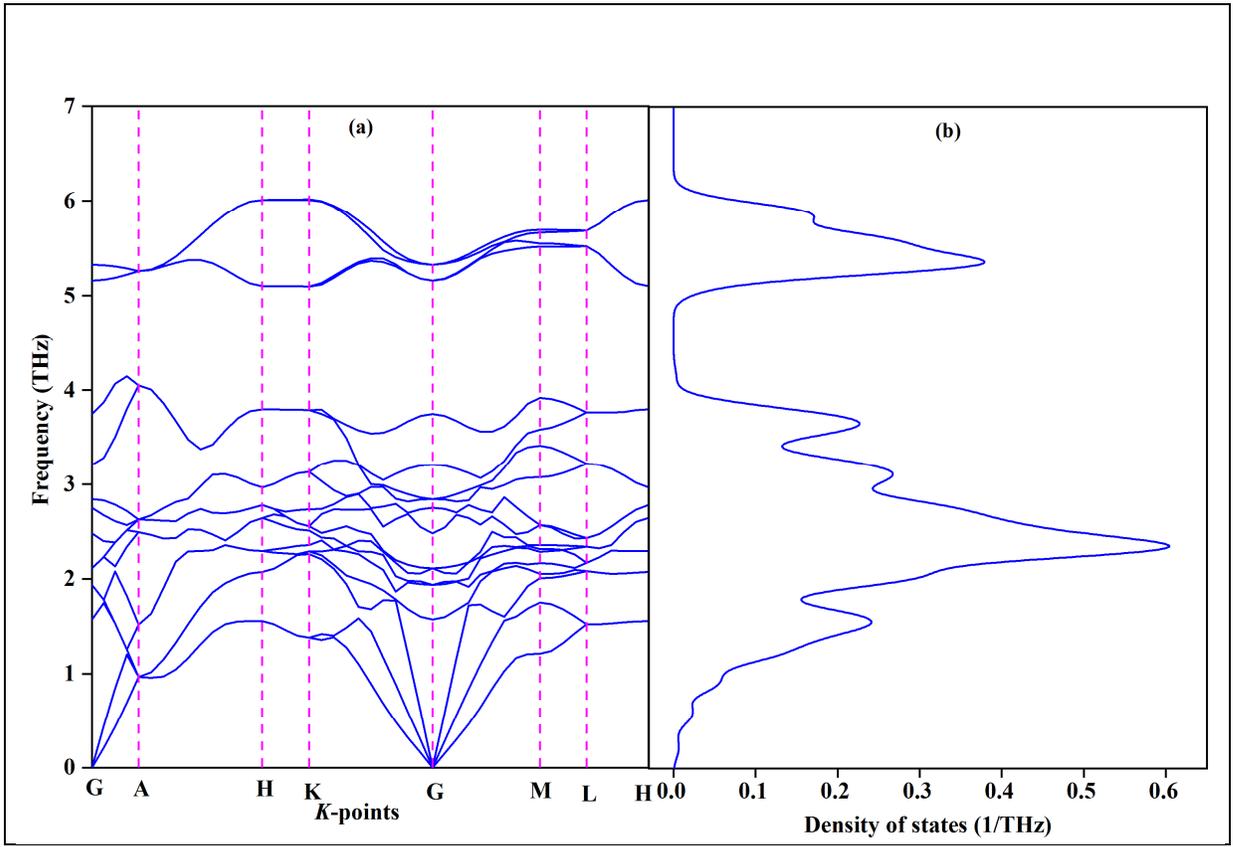

**Figure 6:** Calculated (a) phonon dispersion spectra and the (b) PHDOS for BaAgAs compound at zero pressure.

The acoustic and optical branches of BaAgAs are separated by a frequency gap. The lower branches in phonon dispersion spectra are the acoustic branches and the upper branches represent the optical branches. The acoustic branches are created for in-phase movements of the atoms of the lattice about their equilibrium positions. The acoustic modes at G-point have zero frequency. It is a sign of dynamical stability of the studied compound. The optical properties of crystals are mainly controlled by the optical branches [68]. The high PHDOS regions for the acoustic branches at low phonon frequencies contribute significantly in the thermal transport. We have calculated the total density of phonon states (the right panel of Fig. 6). The PHDOS can be divided into three regions: acoustic modes lower optical modes and upper optical modes. The heavy Ag atoms mainly contribute to the acoustic phonon modes. Lower optical modes contain the contribution of all the atoms. But only the As atom contributes to the upper optical modes. This is expected because As atom is lighter than other atoms. Optical phonons are out-of-phase movements of the atoms in the lattice, one atom moving to the left, and its neighbor to the right. This can only happen if the lattice basis consists of two or more atoms. These modes are called optical because in ionic solids, fluctuations in atomic displacement create an electrical polarization that couples to the electromagnetic field. Thus, these vibrational modes can be excited by infrared radiation. The highest energy phonon dispersion branches of BaAgAs are due to the vibration of light As atoms in the crystal. It is observed that the PHDOS is quite high around the frequency of 5.35 THz in the optical region (Fig. 6). These phonons are expected to play significant role in determining the optical properties of BaAgAs.

### 3.5.2. Acoustic properties

The sound velocity through a material is very important parameter to determine the thermal and electrical behaviors. The average sound velocity in solids, $v_m$, is related to the shear modulus (G) and bulk modulus (B). The $v_m$ is given by the harmonic mean of the average longitudinal and transverse sound velocities, $v_l$ and $v_t$, respectively. The relevant relations are given below [58]:

$$v_m = \left[\frac{1}{3}\left(\frac{1}{v_l^3} + \frac{2}{v_t^3}\right)\right]^{-1/3} \tag{19}$$

$$v_l = \left[\frac{3B + 4G}{3\rho}\right]^{1/2} \tag{20}$$

$$\text{and } v_t = \left[\frac{G}{\rho}\right]^{1/2} \tag{21}$$

Table 7 exhibits the calculated crystal density of and the acoustic velocities in BaAgAs.

**Table 7:** Density ρ (kg/m$^3$), transverse velocity $v_t$ (ms$^{-1}$), longitudinal velocity $v_l$ (ms$^{-1}$) and average elastic wave velocity $v_m$ (ms$^{-1}$) of BaAgAs.

| Compound | ρ | $v_l$ | $v_t$ | $v_m$ | Ref. |
|---|---|---|---|---|---|
| BaAgAs | 3439.15 | 5645.20 | 3267.57 | 3626.32 | This work |
| BaAgP | 5790.09 | 4070.13 | 2460.21 | 2720.76 | [40] |

The sound velocities in BaAgAs are significantly higher than those in BaAgP. The high sound velocity in BaAgAs results from lower crystal density and higher crystal stiffness of this particular Dirac semimetal. Higher sound velocity in BaAgAs implies that the phonon thermal conductivity of this compound should be higher than that of BaAgP.

### 3.6. Thermal properties

#### 3.6.1. Debye temperature

Debye temperature is one of the most prominent thermo-physical parameter of a material. It is related to phonon thermal conductivity, heat capacity, melting temperature, superconducting transition temperature, and electrical conductivity of solids. There are several methods for calculating the Debye temperature, $\theta_D$. Among them, the suggested by Anderson [69] is straightforward and gives reliable estimation of the Debye temperature. The relevant expression is shown below:

$$\theta_D = \frac{h}{k_B}\left[\left(\frac{3n}{4\pi}\right)\frac{N_A \rho}{M}\right]^{1/3} v_m \tag{22}$$

In this equation, h is the Planck's constant and $k_B$ is the Boltzmann constant, n refers to the number of atoms in a molecule, $N_A$ is the Avogadro's number, ρ is the mass density, M is the molecular mass and $v_m$ is the average velocity of sound within the crystalline solid. The calculated Debye temperature is given in Table 8. The Debye temperature of BaAgAs is moderate, 365.53 K. Thus the phonon thermal conductivity of this compound is expected to moderate as well.

### 3.6.2. The melting temperature

The melting temperature ($T_m$) is another necessary parameter for a material which is used to be at elevated temperatures. The melting temperature is related to the bonding strength and cohesive energy of the crystal. These parameters also determine the elastic constants and moduli. Fine et al. [70] developed a formula for calculating the melting temperature of crystals using the single crystal elastic constants as follows:

$$T_m = 354 + 1.5(2C_{11} + C_{33}) \tag{23}$$

The calculated melting temperature of BaAgAs is listed in Table 8. The melting temperature of 864.10 K of BaAgAs is consistent with its moderate Debye temperature, hardness and various elastic moduli.

### 3.6.3. Thermal conductivity

Thermal conductivity is also an important indicator of a material for thermal transport coefficient that gives the measure of the efficiency of heat transfer by a material. This parameter is depends on temperature. In a weak semimetal like BaAgAs, the thermal conductivity should be dominated by the phonon contributions. The minimum phonon thermal conductivity ($K_{min}$) is the limiting value of the thermal conductivity at higher temperature when the phonon contribution to the thermal conductivity ($K_{ph}$) reaches its minimum value and becomes independent of temperature. Based on the Debye model, Clarke deduced the formula to calculate the minimum thermal conductivity ($K_{min}^{Clarke}$) using the following equation [71].

$$K_{min}^{Clarke} = k_B v_m \left[\frac{nrN_A}{M}\right]^{2/3} \tag{24}$$

The minimum phonon thermal conductivity can also be estimated employing the Cahill formalism where the phonon spectrum has been considered within the Einstein model. The Cahill formula [72] for the minimum thermal conductivity is given below:

$$K_{min}^{Cahill} = \frac{k_B}{2.48} n^{\frac{2}{3}} (v_l + 2v_t) \tag{25}$$

The calculated minimum thermal conductivities are tabulated below (Table 8). The minimum thermal conductivity of BaAgAs is very low, comparable to many prospective thermal barrier coating (TBC) materials [73,74].

The temperature dependent phonon thermal conductivity of BaAgAs can also be estimated using the formalism developed by Slack [75]. The Slack formula for temperature dependent phonon thermal conductivity or lattice thermal conductivity is as follows:

$$K_{ph}(T) = A\left[\frac{M_{av} q_D^3 d}{\gamma^2 N^{2/3} T}\right] \tag{26}$$

In this relation, $M_{av}$ is the average atomic mass (Kg/mol) in the molecule, δ denotes the cubic root of average atomic volume, N denotes the number of atoms present in unit cell, and γ denotes the Grüneisen constant calculated using the poison's ratio (σ). A is a parameter (in W-mol/Kg/m$^2$/K$^3$) depending on γ which is calculated from [76]:

$$A(\gamma) = \frac{5.720 \times 10^7 \times 0.849}{2 \times [1 - \frac{0.514}{\gamma} + \frac{0.224}{\gamma^2}]} \tag{27}$$

The anharmonic effect of a solid is determined by the Grüneisen parameter. The Grüneisen parameter γ is an important quantity in the thermodynamics and lattice dynamics because it is related with bulk modulus, heat capacity, thermal expansion coefficient and volume of the solid. A high value of the Grüneisen parameter implies high level of anharmonicity. The Grüneisen parameter can be evaluated from the Poisson's ratio of a solid as follows [77]:

$$\gamma = \frac{3[1+s]}{2[2-3s]} \quad (28)$$

For crystalline materials the values of γ is usually found in the range of 0.80 - 3.50 [77,78]. The calculated value of γ of BaAgAs is 1.49 which is well within the established range. The value is in the medium range which implies the medium level of anharmonic effects in BaAgAs. All the thermo-physical parameters calculated in this section are summarized in Table 8 below. For comparison, some of the relevant results for the isostructural Dirac semimetal BaAgP are also tabulated.

**Table 8:** The Debye temperature $\theta_D$, minimum thermal conductivity $K_{min}$, Grüneisen parameter γ, lattice/phonon thermal conductivity $K_{ph}$ at 300 K and the melting temperature $T_m$ of BaAgAs compound.

| Compound | $\theta_D$ (K) | $K_{min}$ (W/m-K) | | γ | $K_{ph}$ (W/m-K) | $T_m$ (K) | Ref. |
|---|---|---|---|---|---|---|---|
| | | $K_{min}^{Clark}$ | $K_{min}^{Cahill}$ | | | | |
| BaAgAs | 365.53 | 0.709 | 0.776 | 1.49 | 9.27 | 864.10 | This work |
| BaAgP | 255.00 | -- | -- | -- | -- | 807.91* | [40] |

*Calculated from the elastic constants.

Both Debye temperature and the melting point of BaAgP are lower than those of BaAgAs, in complete consistency with the elastic, hardness, and bonding characteristics.

### 3.7. Bond population analysis

The Mulliken bond populations are calculated to explore the bonding nature (ionic, covalent and metallic) of BaAgAs. We have also performed the Hirshfeld population analysis (HPA). The calculated values of atomic populations and the other relevant parameters are given in Table 9. Low value of the charge spilling parameter indicates that the results of Mulliken population analysis (MPA) are of good quality. It is seen from Table 9 that in BaAgAs, electrons are transferred from As and Ba to Ag. It is an indication of ionic bonds. The electron transfer can be attributed to the difference in the electron affinities of As, Ag, and Ba. On the other hand, non-zero effective valence charge implies that there are some covalent contributions as well. The effective valences in the Mulliken population analysis (MPA) and the Hirshfeld population analysis HPA are different. This is expected since MPA depends on the basis sets used to approximate the wave functions of the orbitals while the HPA is independent of the basis sets [28,79,80]. Both MPA and HPA suggest mixed bonding – ionic and covalent, in the BaAgAs.

**Table 9:** Charge Spilling parameter (%), orbital charge (electron), atomic Milliken charge (electron), effective valance (Mulliken & Hirshfeld) (electron) of BaAgAs.

| Charge Spilling (%) | Atomic species | Mulliken atomic population - orbitals | | | | | Mulliken charge | Formal ionic charge | Effective valence (Mulliken) | Hirshfeld charge | Effective valence (Hirshfeld) |
|---|---|---|---|---|---|---|---|---|---|---|---|
| | | s | p | d | f | total | | | | | |
| 0.77 | As | -0.33 | 4.23 | 0.00 | 0.00 | 3.90 | 1.10 | -3 | 1.90 | -0.23 | 2.77 |
| | As | -0.33 | 4.23 | 0.00 | 0.00 | 3.90 | 1.10 | -3 | 1.90 | -0.23 | 2.77 |
| | Ag | 1.54 | 0.43 | 9.84 | 0.00 | 11.81 | -0.81 | +1 | 0.19 | 0.16 | 0.84 |
| | Ag | 1.54 | 0.43 | 9.84 | 0.00 | 11.81 | -0.81 | +1 | 0.19 | 0.16 | 0.84 |
| | Ba | 3.09 | 6.06 | 1.13 | 0.00 | 10.28 | -0.28 | +2 | 1.72 | 0.07 | 1.93 |
| | Ba | 3.09 | 6.06 | 1.13 | 0.00 | 10.28 | -0.28 | +2 | 1.72 | 0.07 | 1.93 |

### 3.8. Optical properties

We have also computed the energy/frequency dependent optical parameters of BaAgAs to explore the possible opportunity of its use in the optoelectronic device sector. In this section, we have calculated the optical properties such as absorption coefficient, dielectric constant, photoconductivity, refractive index, reflectivity and loss function in the photon energy range up to 15 eV with two different electric field polarization directions of [100] and [001]. The methodology for optical calculations is detailed in Refs. [28,81]. A Gaussian smearing of 0.5 eV, a Drude energy 0.05 eV and an unscreened plasma energy of 5 eV were used to calculate the optical parameters as a function of incident photon energy. The Drude term takes care of the intraband electronic transitions due to the absorption of low energy photons. All the computed optical parameters are shown in Fig. 7 below.

The real part of dielectric function, $\varepsilon_1(\omega)$, is shown in Fig. 7a. The spectra of $\varepsilon_1$ start from negative value with a peak around ~1.6 eV and cross the zero line at around 10.8 eV. This is a typical metallic behavior where no band gap exists. Fig. 7a also shows the imaginary part of the dielectric function, $\varepsilon_2(\omega)$. This parameter is related to the photon absorption characteristics of BaAgAs. The position of the peaks and the spectral weight in $\varepsilon_2(\omega)$ are controlled by the electronic energy density of states of the energy levels involved in the optical transition of electrons and the matrix elements of the transition between the two states involved. Sharp peaks in the imaginary part are found at 2.00 eV for [100] and at 2.6 eV for [001] polarization directions of the incident electric field vector. For both polarizations $\varepsilon_2$ gradually decreases with increasing energy and finally goes to zero at ~ 11.2 eV. There is significant optical anisotropy in the real part of the dielectric function with respect to the polarization states of the electric field. The level of anisotropy is much lower in the imaginary part.

The real part of refractive index, $n(\omega)$, and the imaginary part, $k(\omega)$, are shown in Fig. 7b. The low energy value of $n(\omega)$ was found to be high in the infrared and visible region. This real part measures the group velocity of electromagnetic wave in the solid. The imaginary part, known as the extinction coefficient, determines the attenuation of light as it travels through the material. From Fig. 7b, we observe that infrared light is highly attenuated by BaAgAs. Both the real and imaginary parts of the refractive index decrease monotonically at high energies in the ultraviolet (UV) region of the electromagnetic spectrum and finally become almost flat ~11 eV. The optical anisotropy is quite low up to 11 eV.

The variation of absorption coefficients α(ω), as a function of photon energy, are depicted in Fig. 7c. Finite values of α(ω), for both the polarizations at very low energy supports the metallic state of BaAgAs. The absorption coefficient is quite high in the energy range 5 to 10 eV in the UV region. This suggests that BaAgAs is a good absorber of ultraviolet radiation. There is significant optical anisotropy in optical absorption in the energy range from 5 eV to 10 eV.

The photoconductivity is another important parameter for optoelectronic device applications. Optical conductivity as a function of photon energy is presented in Fig. 7d. The low energy photoconductivity reaffirms the metallic character of BaAgAs. There is optical anisotropy in σ(ω). Sharp peaks of real part are found at 2.00 eV for [100] and at 2.5 eV for [001] polarization directions of the incident electric field vector.

The reflectivity, as a function of incident photon energy, is given in Fig. 7e. The reflectivity is higher in the visible region for the [100] polarization. The reflectivity initially decreases in the near-infrared region then increases gradually and becomes almost non-selective in the energy range 3 eV to 11 eV. R(ω) decreases sharply at around 12 eV close to the plasma peak. Reflectivity remains below 40% for electromagnetic radiation with [001] polarization in the visible range. Thus, for this particular polarization, BaAgAs can be used as an anti-reflection material.

The calculated energy loss spectrum is shown in Fig. 7f. The energy loss function helps one to understand the screened plasma excitation created by swift charges inside the material. The loss function, L(ω), shows peak at the characteristic plasma oscillation energy. The position of the peak marks the energy at which the reflectivity and absorption coefficient falls sharply. Above the plasma energy, the system becomes transparent to the incident photons and the optical features become similar to those of insulators. For BaAgAs, the plasma peaks are located at ~11 eV for both electric field polarizations along the [100] and [001] directions.

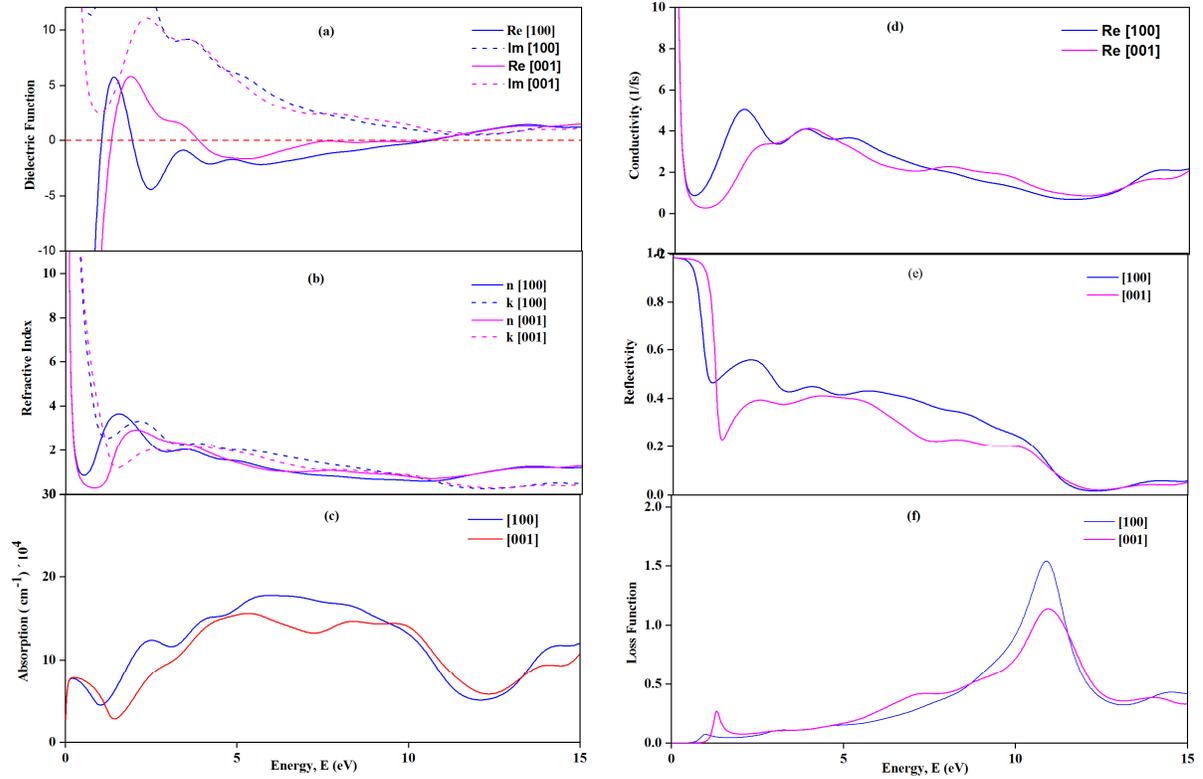

**Figure 7:** The (a) real and imaginary parts of dielectric function [$\varepsilon_1(\omega)$ and $\varepsilon_2(\omega)$], (b) real and imaginary parts of the refractive index [$n(\omega)$ and $k(\omega)$], (c) absorption coefficient [$\alpha(\omega)$], (d) optical conductivity [$\sigma(\omega)$], (e) reflectivity [$R(\omega)$], and (f) loss function [$L(\omega)$] of BaAgAs for the [100] and [001] electric field polarization directions.

## 4. Conclusions

Employing the DFT based first-principles calculations we have explored the elastic, bonding, lattice dynamic, electronic, thermo-physical and optical properties of BaAgAs compound. Most of the reported results are novel. The compound is found to be elastically and stable with brittle features. It is machinable and elastically anisotropic. The phonon dispersion curves confirm dynamical stability of the crystal structure. There are ionic and covalent bondings in BaAgAs. Compared to another isostructural Dirac semimetal BaAgP the bonding strengths and anisotropy level of BaAgAs are higher. The hardness and Debye temperature of BaAsAg are moderate. The phonon thermal conductivity of BaAgAs is low. The electronic band structure shows clear semimetallic character with Dirac point touching the Fermi level. The electronic energy density of states is low. The calculated value of the repulsive Coulomb pseudopotential indicates that BaAgAs has weak electronic correlations. The Fermi surface is an ellipsoid of small size. The optical parameters of BaAgAs have been studied in detail. The compound under study possesses optical anisotropy. BaAgAs is a good absorber of ultraviolet light and reflects visible radiation. The compound is a poor reflector of light. Overall, the optical properties exhibit weak metallic character and are consistent with the electronic band structure calculations.

**Acknowledgements**

S. H. N. acknowledges the research grant (1151/5/52/RU/Science-07/19-20) from the Faculty of Science, University of Rajshahi, Bangladesh, which partly supported this work. A. S. M. M. R. is thankful to the University Grant Commission (UGC), Dhaka, Bangladesh which gave him the Fellowship to conduct this research.

**Data availability**

The data sets generated and/or analyzed in this study are available from the corresponding author on reasonable request.

**Declaration of interest**

The authors declare that they have no known competing financial interests or personal relationships that could have appeared to influence the work reported in this paper.

**CRediT authorship contribution statement**

**A.S.M. Muhasin Reza**: Formal analysis, Methodology, Writing–original draft. **S.H. Naqib**: Supervision, Formal analysis, Conceptualization, Project administration, Writing-review & editing.

**References**

[1]  C.K. Barman, C. Mondal, B. Pathak, A. Alam, Phys. Rev. Mater. 4 (2020) 084201.
[2]  C. Fang, M.J. Gilbert, X. Dai, B.A. Bernevig, Phys. Rev. Lett. 108 (2012) 266802.
[3]  B.J. Wieder, Y. Kim, A.M. Rappe, C.L. Kane, Phys. Rev. Lett. 116 (2016) 186402.
[4]  A.A. Soluyanov, D. Gresch, Z. Wang, Q. Wu, M. Troyer, X. Dai, B.A. Bernevig, Nature 527 (2015) 495–498.
[5]  N.P. Armitage, E.J. Mele, A. Vishwanath, Rev. Mod. Phys. 90 (2018) 015001.
[6]  C. Mondal, C.K. Barman, A. Alam, B. Pathak, Phys. Rev. B 99 (2019) 205112.
[7]  C.K. Barman, C. Mondal, B. Pathak, A. Alam, Phys. Rev. B 99 (2019) 045144.
[8]  Z. Zhu, G.W. Winkler, Q. Wu, J. Li, A.A. Soluyanov, Phys. Rev. X 6 (2016) 031003.
[9]  B. Bradlyn, J. Cano, Z. Wang, M.G. Vergniory, C. Felser, R.J. Cava, B.A. Bernevig, Science 353 (2016) aaf5037.
[10] S. Mardanya, B. Singh, S.-M. Huang, T.-R. Chang, C. Su, H. Lin, A. Agarwal, A. Bansil, Phys. Rev. Mater. 3 (2019) 071201.
[11] F. Merlo, M. Pani, M.L. Fornasini, J. Common Met. 166 (1990) 319–327.
[12] A. Mewis, Z. Für Naturforschung B 34 (1979) 1373–1376.
[13] S. Malick, J. Singh, A. Laha, V. Kanchana, Z. Hossain, D. Kaczorowski, Phys. Rev. B 105 (2022) 045103.
[14] Ph. Barbarat, S.F. Matar, Comput. Mater. Sci. 10 (1998) 368–372.
[15] B. Singh, S. Mardanya, C. Su, H. Lin, A. Agarwal, A. Bansil, Phys. Rev. B 98 (2018) 085122.
[16] K. Peng, Z. Zhou, H. Wang, H. Wu, J. Ying, G. Han, X. Lu, G. Wang, X. Zhou, X. Chen, Adv. Funct. Mater. 31 (2021) 2100583.
[17] H. Nowotny, P. Rogl, J.C. Schuster, J. Solid State Chem. 44 (1982) 126–133.
[18] M. Khazaei, M. Arai, T. Sasaki, M. Estili, Y. Sakka, J. Phys. Condens. Matter 26 (2014) 505503.
[19] M.W. Barsoum, MAX Phases: Properties of Machinable Ternary Carbides and Nitrides, 1. ed, Wiley-VCH, Weinheim, 2013.
[20] S. Aydin, A. Tatar, Y.O. Ciftci, Solid State Sci. 53 (2016) 44–55.
[21] H. Ledbetter, A. Migliori, J. Appl. Phys. 100 (2006) 063516.

[22] B. Cui, W.E. Lee, Refract. Worldforum 5 (2013) 105–112.
[23] S. Li, R. Ahuja, M.W. Barsoum, P. Jena, B. Johansson, Appl. Phys. Lett. 92 (2008) 221907.
[24] Q. Zhu, Q. Wang, L. Li, Z. Yang, J. Yang, B. Chen, C. Cao, H. Wang, J. Du, Phys. Rev. B 104 (2021) 144305.
[25] S. Malick, A. Ghosh, C.K. Barman, A. Alam, Z. Hossain, P. Mandal, J. Nayak, Phys. Rev. B 105 (2022) 165105.
[26] P. Hohenberg, W. Kohn, Phys Rev 136 (1964) B864.
[27] W. Kohn, L.J. Sham, Phys. Rev. 140 (1965) A1133–A1138.
[28] S.J. Clark, M.D. Segall, C.J. Pickard, P.J. Hasnip, M.I.J. Probert, K. Refson, M.C. Payne, Z. Für Krist. - Cryst. Mater. 220 (2005) 567–570.
[29] T.H. Fischer, J. Almlof, J. Phys. Chem. 96 (1992) 9768–9774.
[30] H.J. Monkhorst, J.D. Pack, Phys. Rev. B 13 (1976) 5188–5192.
[31] M.I. Naher, S.H. Naqib, J. Alloys Compd. 829 (2020) 154509.
[32] M.I. Naher, S.H. Naqib, Scientific Reports 11 (2021) 1.
[33] B.R. Rano, I.M. Syed, S.H. Naqib, J. Alloys Compd. 829 (2020) 154522.
[34] B.R. Rano, I.M. Syed, S.H. Naqib, Results in Physics 19 (2020) 103639.
[35] S. Saha, T.P. Sinha, A. Mookerjee, Phys. Rev. B 62 (2000) 8828–8834.
[36] M. Born, Math. Proc. Camb. Philos. Soc. 36 (1940) 160–172.
[37] Félix Mouhat, François-Xavier Coudert, Phys. Rev. B 90 (2014) 224104.
[38] M.A. Ali, M.M. Hossain, M.M. Uddin, A.K.M.A. Islam, D. Jana, S.H. Naqib, J. Alloys Compd. 860 (2021) 158408.
[39] M.A. Hadi, U. Monira, A. Chroneos, S.H. Naqib, A.K.M.A. Islam, N. Kelaidis, R.V. Vovk, J. Phys. Chem. Solids 132 (2019) 38.
[40] F. Parvin, M.A. Hossain, I. Ahmed, K. Akter, A.K.M.A. Islam, Results in Physics 23 (2021) 10468.
[41] W. Voigt, *Lehrbuch der Kristallphysik*, Teubner, Leipzig, (1928).
[42] A. Reuss, ZAMM - J. Appl. Math. Mech. Z. Für Angew. Math. Mech. 9 (1929) 49–58.
[43] R. Hill, Proc. Phys. Soc. Sect. A 65 (1952) 349.
[44] M. Roknuzzaman, M.A. Hadi, M.J. Abden, M.T. Nasir, A.K.M.A. Islam, M.S. Ali, K. Ostrikov, S.H. Naqib, Comput. Mater. Sci. 113 (2016) 148–153.
[45] F. Zerarga, D. Allali, A. Bouhemadou, R. Khenata, B. Deghfel, S. Saad Essaoud, R. Ahmed, Y. Al-Douri, S.S. Safaai, S. Bin-Omran, S.H. Naqib, Computational Condensed Matter 32 (2022) e00705.
[46] D. Vanderbilt, Phys. Rev. B 41 (1990) 7892–7895.
[47] N. Miao, B. Sa, J. Zhou, Z. Sun, Comput. Mater. Sci. 50 (2011) 1559.
[48] X-Q. Chen, H. Niu, D. Li, Y. Li, Intermetallics 19 (2011) 1275.
[49] L. Vitos, P.A. Korzhavyi, B. Johansson, Nature Mater. 2 (2003) 1.
[50] R.C. Lincoln, K.M. Koliwad, P.B. Ghate, Phys Rev 157 (1967) 463.
[51] M.J. Phasha, P.E. Ngoepe, H.R. Chauke, D.G. Pettifor, D. Nguyen-Mann, Intermetallics 18 (2010) 2083.
[52] M.A. Ali, M.M. Hossain, M.A. Hossain, M.T. Nasir, M.M. Uddin, M.Z. Hasan, A.K.M.A. Islam, S.H. Naqib, J. Alloys Compd. 743 (2018) 146.
[53] M.A. Ali, M.A. Hadi, M.M. Hossain, S.H. Naqib, A.K.M.A. Islam, physica status solidi b 254 (2017) 1700010.
[54] H. Mebtouche, O.Baraka, A. Yakoubi, R. Khenata, S.A. Tahir, R. Ahmed, S.H. Naqib, A. Bouhemadou, S. Bin Omran, Xiaotian Wang, Materials Today Communications 25 (2020) 101420.
[55] M.I. Naher, M.A. Afzal, S.H. Naqib, Results in Physics 28 (2021) 104612.
[56] S.F. Pugh, Philos. Mag. J. Sci. 45 (1954) 823–843.

[57] O.L. Anderson, H.H. Demarest Jr., J. Geophys. Res. 76 (1971) 1349.
[58] P. Ravindran, L. Fast, P. Korzhavyi, B. Johansson, J. Wills and O. Eriksson, J. Appl. Phys. 84 (1998) 4891.
[59] Md. Sajidul Islam, Razu Ahmed, Md. Mahamudujjaman, R.S. Islam, S.H. Naqib, Results in Physics 44 (2023) 106214.
[60] D.G. Pettifor, Mater. Sci. Technol. 8 (1992) 345–349.
[61] M.A. Hadi, N. Kelaidis, S.H. Naqib, A. Chroneos, A.K.M.A. Islam, J. Phys. Chem. Solids 129 (2019) 162–171.
[62] M.I. Naher, M. Mahamudujjaman, A. Tasnim, R.S. Islam, S.H. Naqib, Solid State Sciences 131 (2022) 106947.
[63] F. Parvin, S.H. Naqib, J. Alloys Compd. 780 (2019) 452–460.
[64] N.S. Khan, B.R. Rano, I.M. Syed, R.S. Islam, S.H. Naqib, Results in Physics 33 (2022) 105182.
[65] R.C. Lincoln, K.M. Koliwad, P.B. Ghate, Phys. Rev. 157 (1967) 463–466.
[66] K.J. Puttlitz, K.A. Stalter, Handbook of Lead-Free Solder Technology for Microelectronic Assemblies, CRC Press, 2004.
[67] L. Vitos, P.A. Korzhavyi, B. Johansson, Nat. Mater. 2 (2003) 25–28
[68] Y. Yun, D. Legut, P.M. Oppeneer, J. Nucl. Mater. 426 (2012) 109–114.
[69] O.L. Anderson, J. Phys. Chem. Solids 24 (1963) 909–917.
[70] M.E. Fine, L.D. Brown, H.L. Marcus, Scr. Metall. 18 (1984) 951–956.
[71] D. Clarke, C. Levi, Annu. Rev. Mater. Res. 33 (2003) 383–417.
[72] D.G. Cahill, S.K. Watson, R.O. Pohl, Phys. Rev. B 46 (1992) 6131–6140.
[73] M.A. Ali, M.M. Hossain, M.M. Uddin, A.K.M.A. Islam, S.H. Naqib, ACS Omega 8 (2023) 954.
[74] M.A. Hadi, Istiak Ahmed, M.A. Ali, M.M. Hossain, M.T. Nasir, M.L. Ali, S.H. Naqib, A.K.M.A. Islam, Open Ceramics 12 (2022) 100308.
[75] Glen A. Slack, Per Andersson, Phys. Rev. B 26 (1982) 1873.
[76] C.L. Julian, Phys. Rev. 137 (1965) A128–A137.
[77] V.N. Belomestnykh, E.P. Tesleva, Tech. Phys. 49 (2004) 1098–1100.
[78] S.I. Mikitishin, Sov. Mater. Sci. Transl Fiz.-Khimicheskaya Mekhanika Mater. Acad. Sci. Ukr. SSR 18 (1982) 262–265.
[79] R.S. Mulliken, The Journal of Chemical Physics 23 (1955) 1833.
[80] F.L. Hirshfeld, Theoretica chimica acta 44 (1977) 129–138.
[81] M. Mahamudujjaman, M.A. Afzal, R.S. Islam, S.H. Naqib, AIP Advances 12 (2022) 025011.